\documentclass[12pt]{article}
\def\title#1{{\bf\large #1}}

\usepackage[dvips]{graphicx}

\begin{document}

\begin{center}

\vspace*{1.00\baselineskip}

\title{How to apply importance-sampling techniques\\
to simulations of optical systems} \\

\vspace{1.00\baselineskip}

C. J. McKinstrie and P. J. Winzer\\
{\it\small Bell Laboratories, Lucent Technologies, Holmdel, New Jersey 07733} \\

\vspace{0.50\baselineskip}

Abstract \\

\vspace{0.25\baselineskip}

\parbox[t]{5.5in}{\small This report contains a tutorial introduction to the
method of importance sampling. The use of this method is illustrated for
simulations of the noise-induced energy jitter of return-to-zero pulses in
optical communication systems.}

\end{center}

\newpage

\noindent When one rolls a die, the probability of it stopping with a
particular number face-up is 1/6. To measure such a probability experimentally
and accurately, one must roll the die many more than 6 times. With this fact in
mind, consider the design of optical communication systems. A common way to
evaluate system prototypes is to measure (in computer simulations or laboratory
experiments) the distance over which they can transmit information with
bit-error-ratios of order $10^{-9}$ (without forward error correction). Even
with current computers, it is impractical to simulate pulse transmission many
more than $10^9$ times.

One can use importance-sampling techniques to circumvent this difficulty. Put
simply, bit errors are caused by large system perturbations (deviations from
ideal behavior) that occur infrequently. When one makes importance-sampled
simulations of system performance one biases them in such a (controlled) way
that these large perturbations occur more often than they should. Because they
occur often, one can measure their statistics accurately. Subsequently, one
adjusts the simulation results to remove their artificial bias. In this way one
obtains results that are both unbiased and accurate.

Pioneering importance-sampled simulations of phase noise in an optical system
were made by Foschini and Vannuci \cite{fos88}. Recently, importance-sampled
simulations of polarization-mode dispersion were made by Biondini {\it et al.}
\cite{bio02}. Importance-sampling methods were reviewed comprehensively by
Smith {\it et al.} \cite{smi97} and ways to combine data from different
importance-sampled simulations were discussed by Veach \cite{vea97}. In this
report we make no attempt to duplicate their discussions. Rather, we show by
example how to apply importance-sampling techniques to optical systems.

It is instructive to consider the die example quantitatively. Each roll of an
unbiased die is a trial, in which the probability of a successful outcome $p =
1/6$ and the value associated with a successful outcome (the weight with which
a successful outcome is counted) $w = 1$. For each trial the expected value of
the outcome is $pw$ and the variance is $w^2p(1 - p)$. Suppose that an unbiased
die is rolled $n_1$ times and let $f_1$ be the measured 1-frequency (the number
of 1s divided by the number of rolls). Then the expected value of the
1-frequency $E(f_1) = 1/6$. By adding $n_1$ terms of magnitude $5/36$ and
dividing the result by $n_1^2$ one finds that the variance $V(f_1) = 5/36n_1$:
To measure the 1-frequency with an accuracy of 1\% (to conduct an experiment
for which the standard deviation of the 1-frequency is 1\% of the expected
value) one would have to roll the die $5 \times 10^4$ times. This would be a
time-consuming task.

If the 5- and 6-frequencies were of no interest one could bias the die by
marking the 5- and 6-faces with 1s. Since this change would increase the
probability of rolling a 1 ($p = 3/6 = 1/2$), one should count each successful
roll with reduced weight ($w = 1/3$). Suppose that this biased die is rolled
$n_2$ times and let $f_2$ be the measured (weighted) 1-frequency. Then the
expected value $E(f_2) = 1/6$, as required, and the variance $V(f_2) =
1/36n_2$: By causing the desired event to occur more often than it would
naturally, one is able to measure its probability more accurately, or with
fewer rolls. (For rare events that occur in applications, the performance
improvements are much larger than the factor of 5 associated with this
example.) Of course, the price one pays for this increase in accuracy is the
loss of information about the 5- and 6-frequencies.

If these frequencies were of limited, but finite, interest, one could combine
the measurements made with each die separately. If one were to weight the two
measurements equally, by defining the combined 1-frequency $f = (f_1 + f_2)/2$,
one would find that $E(f) = 1/6$ and $V(f) = 5/144n_1 + 1/144n_2$: The accuracy
of the combined measurement is limited by the less-accurate individual
measurement (the one with the larger variance coefficient or the one made with
fewer rolls). One should weight the individual measurements in proportion to
the numbers of rolls used to make them, and in inverse proportion to their
variance coefficients. Let $f = c_1f_1 + c_2f_2$, where the condition $c_1 +
c_2 = 1$ ensures that $E(f) = 1/6$. Then a short calculation shows that the
optimal values of $c_1$ and $c_2$ are $n_1/(n_1 + 5n_2)$ and $5n_2/(n_1 +
5n_2)$, respectively, in which case the minimal variance $V(f) = 5/36(n_1 +
5n_2)$. For the case in which $n_1 = n_2 = n/2$, where $n$ is the total number
of rolls, $V(f) = 5/108n$.

In this dice example simple formulas for the individual variances exist, which
allow one to determine the optimal weight coefficients precisely. However, in
applications such formulas might be complicated or unknown. Consequently, a
different method is required. An effective method, which is called the balance
heuristic \cite{vea97}, is to weight each successful outcome (roll of either
die) equally. In this method, if the unbiased die is rolled $n_1$ times and the
biased die is rolled $n_2$ times, the combined probability of rolling a 1 is
$(n_1 + 3n_2)/6(n_1 + n_2)$. To ensure that the expected value of the combined
measurement is 1/6, each successful roll (of either die) should be counted with
weight $w = (n_1 + n_2)/ (n_1 + 3n_2)$. By adding $n_1$ contributions of
magnitude $w^2p_1(1 - p_1)$ and $n_2$ contributions of magnitude $w^2p_2(1 -
p_2)$, and dividing the result by $(n_1 + n_2)^2$, one finds that the combined
variance is $(5n_1 + 9n_2)/36(n_1 + 3n_2)^2$. For the case in which $n_1 = n_2
= n/2$ the combined variance is $7/144n$, which is only 5\% larger than the
minimal variance of the preceding paragraph. Thus, one should weight each
outcome in inverse proportion to its combined probability.

Now consider the noise-induced energy (amplitude) jitter of a return-to-zero
(RZ) pulse. Provided that the pulse energy $e$ is greater than the noise energy
in the surrounding bit slot, it evolves according to the stochastic ordinary
differential equation (ODE)
\begin{equation}
d_ze = (g - \alpha)e + r, \label{1}
\end{equation}
where $g(z)$ is the amplifier gain rate, $\alpha$ is the fiber loss rate and
$r(z)$ is the rate at which the energy is changed by amplifier noise
\cite{mck02,mck03a}. This random rate of change is quantified by the equations
$\langle r(z)\rangle = 0$ and $\langle r(z)r(z')\rangle = (2n_{{\rm sp}}h\nu
ge)\delta(z - z')$, where $\langle\ \rangle$ denotes an ensemble average,
$n_{{\rm sp}}$ is the spontaneous-emission factor (1.1--1.3) and $h\nu$ is the
photon energy. Equation (1) is valid for any isolated pulse and an arbitrary
combination of distributed and lumped amplification.

For definiteness, consider a 10 Gb/s system with uniformly-distributed
amplification ($g = \alpha$), in which $\alpha = 0.21$ dB/Km, $\beta = -0.30$
ps$^2$/Km (D = 0.38 ps/Km-nm) and $\gamma =$ 1.7/Km-W. Then a soliton with a
full-width at half-maximum of 30 ps has an energy of 21 fJ (time-averaged power
of 0.21 mW). If the system length $l = 10$ Mm the output noise power in both
polarizations, in a frequency bandwidth of 12 GHz (wavelength bandwidth of 0.1
nm), is 1.7 $\mu$W: The (optical) signal-to-noise ratio is 21 dB. Systems with
nonuniformly-distributed or lumped amplification produce the same noise power
in shorter distances.

For uniformly-distributed amplification Eq. (\ref{1}) can be rewritten in the
canonical form
\begin{equation}
dx = x^{1/2}dy, \label{2}
\end{equation}
where $x = e/e_0$ is the energy, normalized to the equilibrium energy (in the
absence of noise), and $y$ is a Wiener process (Gaussian random variable) with
$\langle y\rangle = 0$ and $\langle y^2\rangle = \sigma_s^2z$, where the
normalized source-strength $\sigma_s^2 = 2n_{{\rm sp}}h\nu g/e_0$. In the
linear regime the multiplicative factor $x^{1/2} \approx 1$, from which it
follows that $x \approx 1 + y$: The probability-density function (PDF) of the
output energies is Gaussian, with mean 1 and variance $\sigma_s^2 z$. (From a
logical standpoint the PDF of the non-negative quantity $x$ cannot be exactly
Gaussian, because, if it were, the probability of $x < 0$ would be finite for
all $z > 0$. From a practical standpoint this inconsistency is tolerable if the
probability of $x < 0$ is exponentially small for system lengths of interest.)
For the aforementioned system the output variance $\sigma_s^2l = 6.6 \times
10^{-3}$ (which is of order $10^{-2}$) and the output deviation is $8.1 \times
10^{-2}$ (which is of order $10^{-1}$). In the nonlinear regime the factor
$x^{1/2}$ modifies the tails of the PDF significantly. For reference, the
analytical solution of Eq. (\ref{2}) has the PDF
\begin{equation}
P(x) \approx {\cosh(mx^{1/2}/v)\exp[-(m^2 + x)/2v] \over (2\pi xv)^{1/2}},
\label{3}
\end{equation}
where $m = 1 - \sigma_s^2z/8$ and $v = \sigma_s^2z/4$ \cite{mck03b}.

Equation (2) and solution (3) model energy jitter in a (continuous) system with
uniformly-distributed amplification. We simulated a (discrete) system with $n_i
= 100$ lumped amplifiers. Between the amplifiers the energy $x$ did not change.
At the $i$th amplifier the energy was changed (kicked) by the random amount
$\delta y_i$, where the properties $\langle\delta y_i\rangle = 0$ and $\langle
\delta y_i\delta y_j\rangle = 10^{-4}\delta_{ij}$ ensured that
$\big\langle(\sum_{i=1}^{n_i}\delta y_i)^2\big\rangle = 10^{-2}$: The discrete
system had the same characteristics as the continuous system. The output
energies were assigned to energy bins of (common) width 0.02 and each bin
probability $p_j$ was defined to be the number of pulses whose energies fell
within the bin boundaries (bin count) divided by the total number of pulses.
(Since probabilities cannot be measured by finite numbers of trials, these
quantities should be called the relative frequencies associated with the bins.
We use the term probabilities as an abbreviation for the correct term.) To
facilitate comparisons to the analytical PDF (\ref{3}), the simulation
probabilities were defined to be the bin probabilities divided by the bin
width. In these (direct) simulations the occurrence of each output energy was
counted with unit weight.

The PDF associated with an ensemble of $10^6$ pulses is displayed in Fig. 1.
Although the simulation results agree well with Eq. (\ref{3}) near the peak of
the PDF, they do not even begin to sample the tail of the PDF. On a 1-GHz PC
these simulations, which are based on Eq.~(2), take a few minutes. Simulating
the transmission of many more than $10^9$ pulses would take many days.
Realistic simulations, which are based on the nonlinear Schroedinger equation,
would take even longer.

To probe the tails of the PDF one must make large energy perturbations occur
more often than they would naturally. One way to achieve this goal is to
increase the standard deviation of the energy kicks. Let $q_0$ denote the
(common) unbiased kick distribution, with deviation $\sigma_0 = 10^{-2}$, and
$q$ denote the (common) biased kick distribution, with deviation $\sigma >
\sigma_0$. Then, at the $i$th amplifier the probability that the energy is
kicked by the amount $\delta y_i$ is increased by the factor $q(\delta
y_i)/q_0(\delta y_i)$. Since the kicks at all the amplifiers are biased, the
output energy occurs with a probability that is larger than its natural
probability by the total factor $f_t = \Pi_{i=1}^{n_i}q(\delta y_i)/q_0(\delta
y_i)$, which depends on the full kick sequence $(\delta y_1,\delta
y_2,\dots,\delta y_{n_i})$. One can remove this bias by counting the output
energy with reduced weight: One increments the appropriate bin probability by
$1/f_t$, rather than~1. All other aspects of data counting remain the same. For
reference, the probability factor $1/f_t$ is called the likelihood ratio.

The PDF associated with an ensemble of $10^6$ pulses is displayed in Fig. 2 for
the case in which $\sigma = 1.2\sigma_0$. The results of these
(importance-sampled) simulations differ from the previous results in two ways.
First, they do probe the tail of the PDF. Although the simulations reproduce
the shape of the analytical PDF, the simulation probabilities are not accurate
because the number of data points that sample the tail of the PDF is still
small. Second, by causing large kicks to happen more often, one causes small
kicks to happen less often. Consequently, the body (peak) of the PDF is not
reproduced accurately. This deficiency prevents one from increasing $\sigma$
until the tail of the PDF is sampled accurately.

Another way to achieve the stated goal is to change the (common) mean of the
kick distributions. If the mean kick $\mu$ is positive (negative) the energy
drifts toward larger (smaller) values. The distributions associated with 3
ensembles of $3 \times 10^5$ pulses are displayed in Fig. 3 for cases in which
$\sigma = \sigma_0$, and $\mu = -3 \times 10^{-3}$, 0 and $3 \times 10^{-3}$.
These values produce data sets with mean energies of $-0.3$, 0.0 and 0.3,
respectively. Although the simulation distributions are inaccurate for energies
that are far from their mean energies (because the bin counts are low), they
are accurate near their mean energies. It only remains to combine the
individual distributions to produce a composite distribution that is accurate
for the entire domain of interest.

One way to combine the individual distributions is to weight their bin
probabilities equally (which is equivalent to combining the data sets before
sorting the output energies and the associated probability factors into bins).
Let $p_{jk}$ be the $j$th bin probability associated with the $k$th data set
(which was produced by kick distributions with mean $\mu_k$). For each $j$, if
the individual bin probabilities $p_{jk}$ were all zero the combined bin
probability $p_j$ was defined to be zero and if some of the $p_{jk}$ were
nonzero $p_j$ was defined to be their average. The results of this procedure
are shown in Fig. 4. Although the composite distribution does cover the domain
of interest, it is inaccurate near the boundaries of the sample spaces.
Combining the individual distributions with equal weight allows the bodies of
the distributions (which have high bin counts) to be polluted by the tails of
neighboring distributions (which have low bin counts).

The dice example suggests that it is better to weight the bin probabilities
according to the associated bin counts. For each $j$, if the individual bin
counts $b_{jk}$ were all zero the combined bin probability $p_j$ was defined to
be zero and if some of the bin counts were nonzero $p_j$ was defined to be
$\sum_{k=1}^{n_k}b_{jk}p_{jk}/\sum_{k=1}^{n_k}b_{jk}$. The results of this
procedure are shown in Fig. 5. The composite distribution covers, and is
accurate throughout, the domain of interest. This {\it ad-hoc} procedure allows
one to combine bin probabilities generated at different times, without recourse
to the data sets on which they were based.

Although the {\it ad-hoc} method works, it is not the balance heuristic.
Suppose that the kick sequence $(\delta y_1,\delta y_2,\dots,\delta y_{n_i})$
occurs during the first simulation, which is made using the biased distribution
$q_1$ (with mean $\mu_1$). Then the associated probability factor $f_{t1} =
\Pi_{i=1}^{n_i}q_1(\delta y_i)/q_0(\delta y_i)$. Were the same sequence to
occur during the $k$th simulation, the associated probability factor would be
$f_{tk}$, which depends on the biased distribution $q_k$. If the simulations
involve the same number of pulses, the combined probability factor $f_t =
\sum_{k=1}^{n_k}f_{tk}/n_k$. (It is easy to generalize this formula.) When we
made the individual simulations and sorted the output energies and the
individual probability factors into bins, we also sorted the combined
probability factors into a separate set of bins, which was common to all the
simulations. The results of this procedure are shown in Fig. 6. The
balance-heuristic method works well.

In summary, we showed by example how to apply importance sampling techniques to
simulations of optical communication systems. These techniques are easy to
apply and increase significantly the accuracy with which rare events can be
simulated.

We acknowledge useful discussions with D. Chizhik, G. Foschini, R. Moore and J.
Salz.

\vspace{\baselineskip}

\noindent Postscript: Independent simulations of energy jitter were made
recently by Moore {\it et al.} [Opt. Lett. {\bf 28}, 105 (2003)], who showed
that the predictions of the energy equation (\ref{1}) are consistent with the
results of simulations based on the nonlinear Schroedinger equation.


\newpage

\begin{figure}[!h]
\centerline{\includegraphics{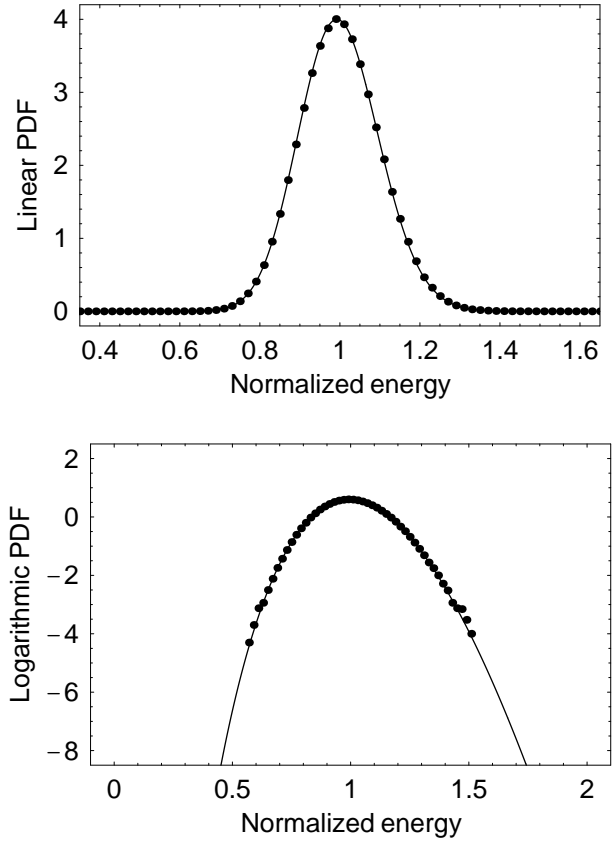}} \caption{Probability distribution
function of the (normalized) output energies obtained by solving Eq.
(\protect\ref{2}) analytically (curve) and numerically, for an ensemble of
$10^6$ pulses (dots). The standard deviation of the energy kicks was
$10^{-2}$.}
\end{figure}

\newpage

\begin{figure}[!h]
\centerline{\includegraphics{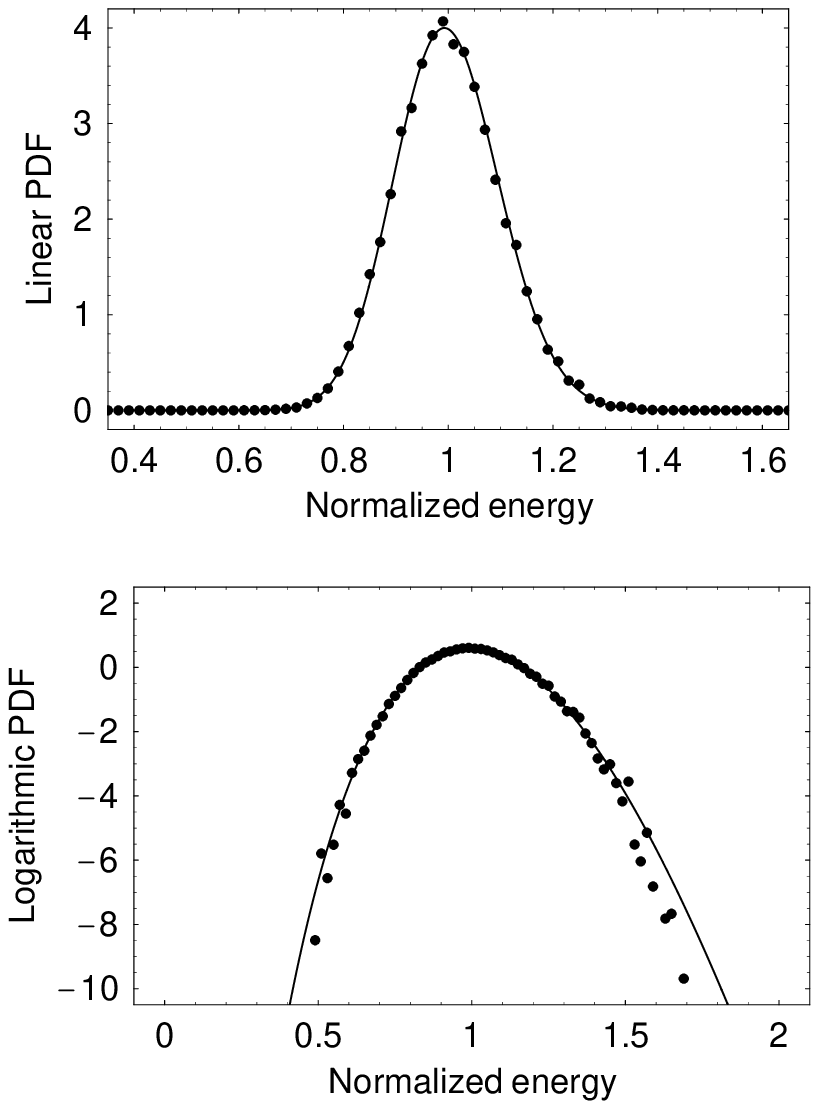}} \caption{Probability distribution
function of the (normalized) output energies obtained by solving Eq.
(\protect\ref{2}) analytically (curve) and numerically, for an ensemble of
$10^6$ pulses (dots). The standard deviation of the energy kicks was $1.2
\times 10^{-2}$.}
\end{figure}

\newpage

\begin{figure}[!h]
\centerline{\includegraphics{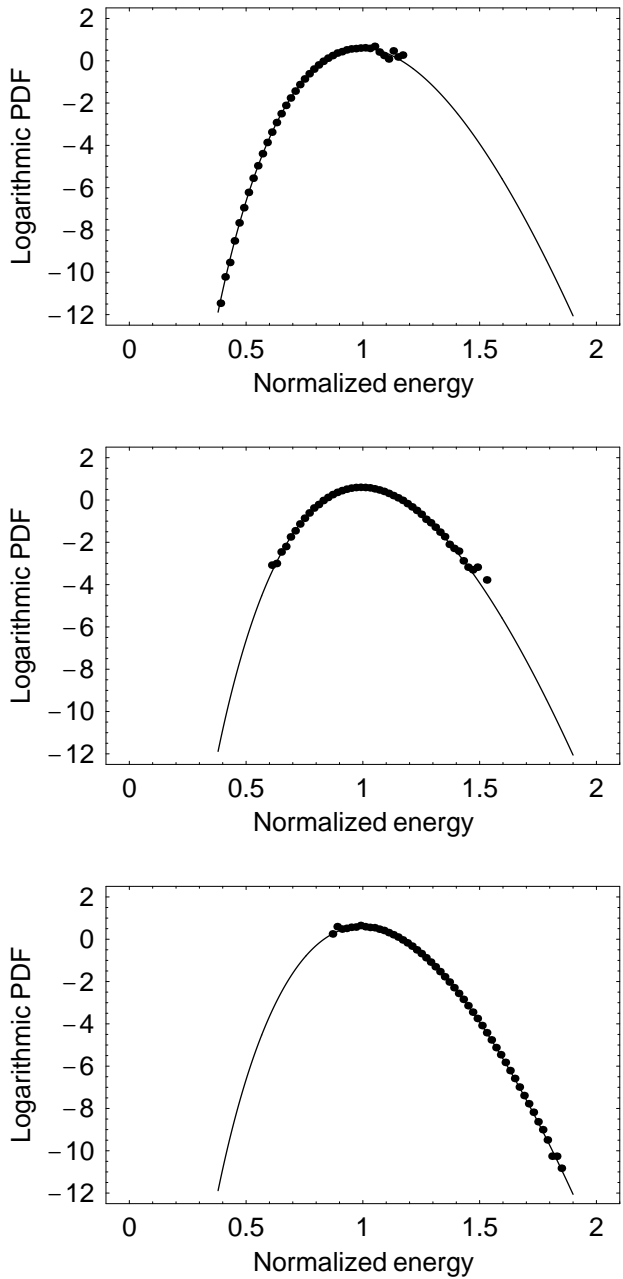}} \caption{Probability distribution
functions of the (normalized) output energies obtained by solving Eq.
(\protect\ref{2}) analytically (curve) and numerically, for 3 ensembles of $3
\times 10^5$ pulses (dots). For each ensemble the standard deviation of the
energy kicks was $10^{-2}$. The mean energy kicks were $-3 \times 10^{-3}$, 0
and $3 \times 10^{-3}$.}
\end{figure}

\newpage

\begin{figure}[!h]
\centerline{\includegraphics{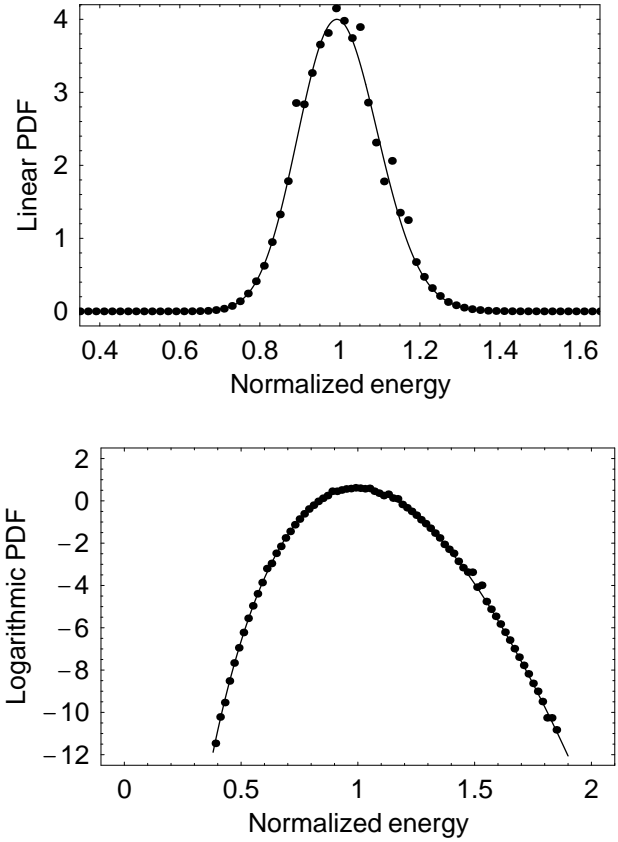}} \caption{Probability distribution
function of the (normalized) output energies obtained by solving Eq.
(\protect\ref{2}) analytically (curve) and numerically, for 3 ensembles of $3
\times 10^5$ pulses (dots). For each ensemble the standard deviation of the
energy kicks was $10^{-2}$. The mean energy kicks were $-3 \times 10^{-3}$, 0
and $3 \times 10^{-3}$. The 3 sets of bin probabilities were combined without
weighting.}
\end{figure}

\newpage

\begin{figure}[!h]
\centerline{\includegraphics{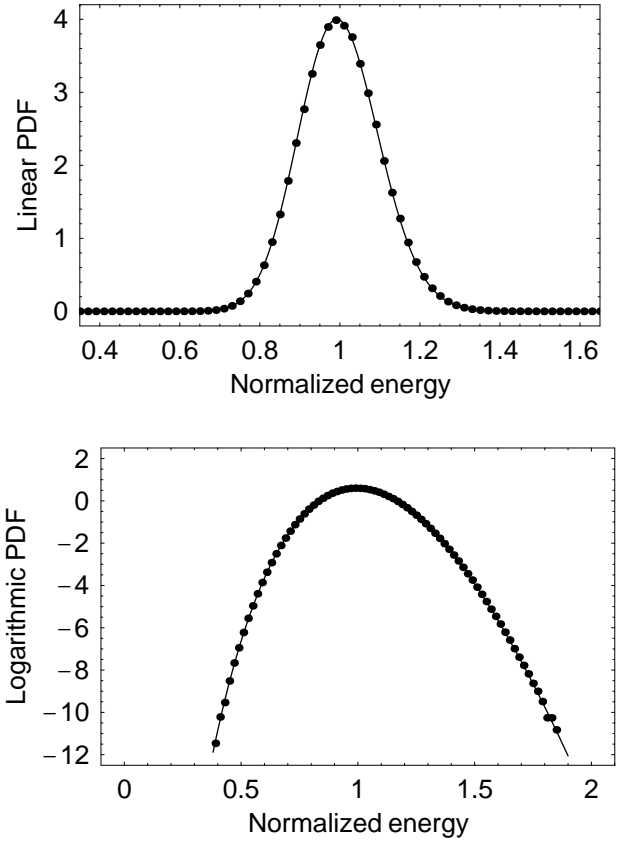}} \caption{Probability distribution
function of the (normalized) output energies obtained by solving Eq.
(\protect\ref{2}) analytically (curve) and numerically, for 3 ensembles of $3
\times 10^5$ pulses (dots). For each ensemble the standard deviation of the
energy kicks was $10^{-2}$. The mean energy kicks were $-3 \times 10^{-3}$, 0
and $3 \times 10^{-3}$. When the 3 sets of bin probabilities were combined,
they were weighted according to the associated bin counts.}
\end{figure}

\newpage

\begin{figure}[!h]
\centerline{\includegraphics{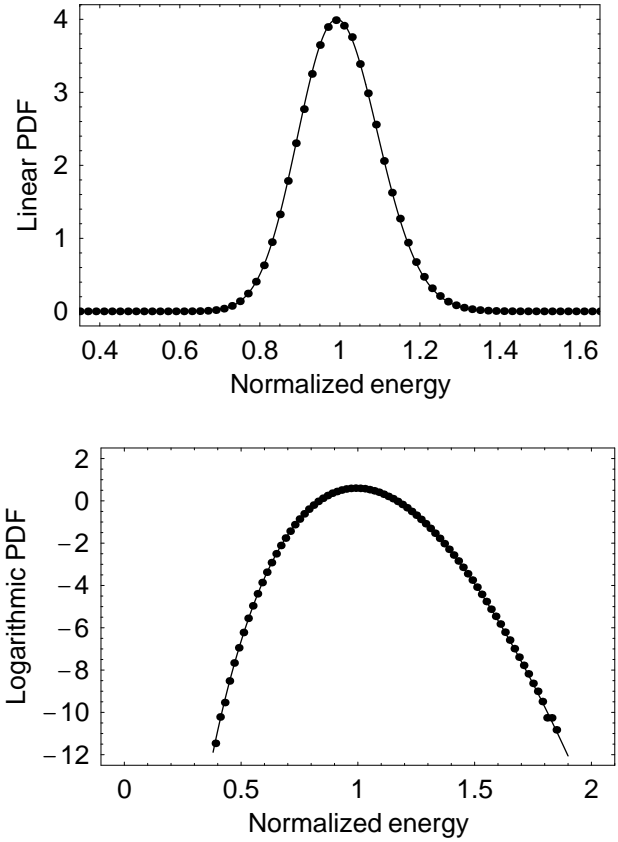}} \caption{Probability distribution
function of the (normalized) output energies obtained by solving Eq.
(\protect\ref{2}) analytically (curve) and numerically, for 3 ensembles of $3
\times 10^5$ pulses (dots). For each ensemble the standard deviation of the
energy kicks was $10^{-2}$. The mean energy kicks were $-3 \times 10^{-3}$, 0
and $3 \times 10^{-3}$. When the 3 data sets were combined, the data were
weighted according to the combined probability factors.}
\end{figure}

\end{document}